%Paper: astro-ph/9510058
%From: metzler@hydra.fnal.gov (Chris Metzler)
%Date: Tue, 10 Oct 95 17:43:59 -0500

\magnification=1200
\parindent=0.0truept
\parskip=6.0truept
\baselineskip=16.0 truept
% \baselineskip=24.0 truept		%	(APJ)
%

\def \mpc       {{\rm\ Mpc}}
\def \kpc       {{\rm\ kpc}}

\def \kms       {\hbox{ km s$^{-1}$}}

\def \ie        {\hbox{\it i.e.}}
\def \eg        {\hbox{\it e.g.}}

\def \etal      {{\it et al.\ }}

\def \dlog      {\hbox{\rm d$\,$log }}

\def \Ho        {{\rm\ H_{0}}}
\def \kmsmpc    {{\rm\ km\ s^{-1}\ Mpc^{-1}}}
\def \kev       {{\rm\ keV}}
\def \msol      {{\rm M}_\odot}
\def \hinv      {\hbox{$\, h^{-1}$} }

\def \arcmin    {^\prime}
\def \cntflux   {{\rm\ cnts\ s^{-1}\ arcmin^{-2}}}

\def \se        {\!=\!}
\def \sims    	{\sim \!}
\def \ssim      {\! \sim \!}
\def \ssimeq    {\! \simeq \!}
\def \sequiv    {\! \equiv \!}

\def\\{\hfil\break}
\def\spose#1{\hbox to 0pt{#1\hss}}
\def\lta{\mathrel{\spose{\lower 3pt\hbox{$\mathchar"218$}}
     \raise 2.0pt\hbox{$\mathchar"13C$}}}
\def\gta{\mathrel{\spose{\lower 3pt\hbox{$\mathchar"218$}}
     \raise 2.0pt\hbox{$\mathchar"13E$}}}

\newcount\itemno
\itemno=1
\def \ino         { \the\itemno\global\advance\itemno by 1 }
\def \bref{\par \noindent \hangindent=1.5 truecm \hangafter=1}  % References
\def\apj{ApJ}

\def\apjs{ApJS}
\def\aa{A\&A}
\def\aj{AJ}
\def\mnras{MNRAS}

\def\nature{Nature}

%
%----- Make comments stand out

%
%----- number only pages greater than one
\footline={\ifnum\pageno>1{\hss\tenrm\folio\hss}\else{\hfil}\fi}
%
%----- Set headline string
\def\today{\ifcase\month\or January\or February\or March\or April\or
    May\or June\or July\or August\or September\or October\or
    November\or December\fi \space\number\day, \number\year}
%
% \headline={\hfil{\it Draft Version --- Not for Distribution --- \today}\hfil}
%
%
%%%%%%%%%%%%%%%%%%%  PAPER BEGINS HERE  %%%%%%%%%%%%%%%%%%%
%
%  Paper specific macros
\def \xray {\hbox{X--ray} }
\def \rfiveh {\hbox{$r_{500}$} }
\def \rfivest {\hbox{$r_{500}$} }
\def \rhocrit {\hbox{$\rho_{c}$} }
\null
% \hskip 4.4truein
% FERMILAB--Pub--95/337--A
\vskip 0.5truein

\centerline {\bf MASS ESTIMATES OF X--RAY CLUSTERS}
\bigskip
\centerline { August E. Evrard\footnote{$^\dagger$} {\sl also Institute for
Theoretical Physics, University of California, Santa Barbara, CA
93106} }
\smallskip
\centerline { Physics Department, University of Michigan, Ann Arbor, MI
48109-1120 USA}
\smallskip
\centerline {\sl evrard@umich.edu }
\medskip
\centerline {Christopher A. Metzler}
\smallskip
\centerline { NASA/Fermilab Astrophysics Center,
% Fermi National Accelerator Laboratory,
Batavia, IL 60510-0500 USA }
\smallskip
\centerline {\sl metzler@umich.edu }
\medskip
\centerline {and}
\medskip
\centerline { Julio F. Navarro$^\dagger$}
\smallskip
\centerline {Steward Observatory, University of Arizona, Tucson, AZ 85721 USA}
\smallskip
\centerline {\sl jnavarro@as.arizona.edu}

\bigskip
\bigskip

\centerline { Abstract }

We use cosmological gas dynamic simulations to investigate the
accuracy of galaxy cluster mass estimates based on \xray observations.
The experiments follow the formation of clusters in different
cosmological models and include the effects of gravity, pressure
gradients, and hydrodynamical shocks. A subset of our ensemble also
allows for feedback of mass and energy from galactic winds into the
intracluster medium.  We find that mass estimates based on the
hydrostatic, isothermal $\beta$-model are remarkably accurate when
evaluated at radii where the cluster mean density is between
$500$-$2500$ times the critical density.  Applied to 174 artificial
ROSAT images constructed from the simulations, the distribution of the
estimated-to-true mass ratio is nearly unbiased and has a standard
deviation of $14$-$29\%$.  The scatter can be considerably reduced (to
$8$-$15\%$) by using an alternative mass estimator that exploits the
tightness of the mass-temperature relation found in the simulations.
The improvement over $\beta$-model estimates is due to the elimination
of the variance contributed by the gas outer slope parameter.  We
discuss these findings and their implications for recent measurements
of cluster baryon fractions.

\vfil
% {\sl submitted to the Astrophysical Journal}

\vfill \eject

{\bf 1. Introduction }
\smallskip

Estimates of the total masses of groups and clusters of galaxies have
been used to infer the amount of dark matter in the universe for over
sixty years.   A novel technique for estimating the density parameter
$\Omega_0$ makes use of precise measurements of the visible, baryonic
mass fraction $f_b$ in galaxy clusters along with limits on the
universal baryon fraction $\Omega_b$ derived from primordial
nucleosynthesis.  White \etal (1993) showed that cluster baryon
fractions, defined as the ratio of the mass in galaxies and
intracluster gas to the total cluster mass, should not differ
substantially from the universal value, $\Omega_b/\Omega_0$, when
determined near the outer boundary of their hydrostatic regions ---
roughly an Abell radius for a cluster as rich as Coma.  A
straightforward and unbiased estimate of the density parameter is
formed by $\Omega_0 \se \Omega_b/f_b$.

Recent measurements in rich clusters (Briel, Henry \& Bohringer 1992;
Durret \etal 1994; David, Jones \& Forman 1995; White \& Fabian 1995)
and poor clusters and groups (Ponman \etal 1994; Pildis, Bregman \&
Evrard 1995; Neumann \& Bohringer 1995) indicate $f_b \!  \ge \! 0.04
\, h^{-3/2}$.  (Hereafter we write Hubble's constant as $100 \, h$ km
s$^{-1}$ Mpc$^{-1}$.)   Taken with the nucleosynthesis determination
$\Omega_b \ssimeq 0.0125 h^{-2}$ (Walker \etal 1991), this implies a
rather small value of the density parameter, $\Omega_0 < 0.3 h^{-1/2}$,
unless primordial nucleosynthesis calculations have underestimated the
universal baryon fraction by almost a factor of three.    Although
there is current debate about the uncertainty in primordial
nucleosynthesis determinations of $\Omega_b$ (Krauss \& Kernan 1994;
Copi, Schramm \& Turner 1995; Hata \etal 1995; Steigman 1995; Sasselov
\& Goldwirth 1995), current interpretation of the data appear to rule
out the large values of $\Omega_b$ required for consistency with a
universe with closure density.

These upper limits on $\Omega_0$ are especially strong because most of
the baryons in clusters are in the hot intracluster medium (ICM), a
component empirically found to be more extended than the dark matter
distribution (\eg, David \etal 1995). Numerical simulations show that
this is a general result of hierarchical scenarios, where clusters are
formed through mergers of protoclusters during which energy is
transferred systematically to the gas from dark matter (Navarro \&
White 1993; Pearce, Thomas \& Couchman 1994).  The results imply that
cluster baryon fractions, measured at radii encompassing a density
contrast of a few hundred, should be about $10-20 \%$ {\it lower} than
the universal value (Evrard 1990; Cen \& Ostriker 1992).

Several solutions have been proposed to rescue $\Omega \se 1$ from what
has been deemed the cluster ``baryon catastrophe'' (Carr 1993).  A low
Hubble constant H$_0 \lta 30 \kmsmpc$ can alleviate the problem
(Bartlett \etal 1995), but is in strong disagreement with recent
observational estimates which favour a high value for H$_0$ (Freedman
\etal 1994; Mould \etal 1995).  A simple possibility which has not yet
been explored in detail is that the binding masses inferred from \xray
observations may be systematically underestimated by a significant
amount. These estimates usually rely on assumptions such as spherical
symmetry, hydrostatic equilibrium for the gas, and virial equilibrium
for the galaxies, none of which may be fully realized in practice.
Clusters exhibit signatures of substructure both in the ICM (Mohr \etal
1995; Buote \& Tsai 1995) and in their galaxy distributions (Dressler
\& Schectman 1988; Bird 1995; Crone, Evrard \& Richstone 1995),
suggestive of recent mass accretion and of significant departures from
equilibrium.  A moderate bias in the mass estimates introduced by these
effects would have important consequences for the $\Omega_0$ limits
derived by the above arguments.

Discrepancies of $\sim 50\%$ have, in fact, been reported when mass
estimates of the central regions of clusters derived from X-ray
observations are compared with those required to produce strong arcs by
gravitational lensing of background galaxies (Miralde-Escud\'e \& Babul
1995).  Although this discrepancy could signal significant departures
from hydrostatic equilibrium or support from non--thermal sources such
as magnetic fields (Loeb \& Mao 1994), systematic errors in the lensing
mass estimates due to projection effects and substructure are more
likely to be responsible for the disagreement (Bartelmann, Steinmetz \&
Weiss 1995; Bartelmann 1995). In fact, at larger radii, weak
gravitational lensing has also been used to measure cluster masses
(Tyson, Valdes \& Wenk 1990; Bonnet, Mellier \& Fort 1994;  Fahlman
\etal 1994) and the small number of clusters with both \xray and weak
lensing mass estimates show no significant discrepancy between the two
(Smail \etal 1995; Squires \etal 1995), although the statistical
uncertainties are still large.

In this paper, we examine the accuracy of \xray binding mass estimates
using a large number of high resolution simulations designed to follow
the non--linear, dynamical evolution of the gravitationally coupled
system of dark matter, gas and (in one set of runs) galaxies in a
variety of cosmological models.  A total of 58 clusters are drawn from
three separate projects using two different Lagrangian hydrodynamical
codes. The models are used to generate 174 synthetic ROSAT \xray images
and broad beam temperature estimates.  Binding masses are then
estimated for these systems in a manner analogous to that applied to
observational datasets. These models are therefore ideal for
understanding possible systematic effects afflicting cluster mass
estimates based on X-ray observations. A recent paper by Schindler
(1995) addresses the same issue using a sample of six simulations
generated with very different techniques.  Our results are in excellent
agreement, despite the fact that the numerical methods and data
analysis procedures used in the two studies differ in a number of
details.

After describing briefly the numerical simulations in \S2, we begin by
examining the validity of the hydrostatic and isothermal assumptions
using the three dimensional velocity and temperature profiles of
simulated clusters (\S3). We then investigate the accuracy of binding
masses estimated using the simplest combination of \xray imaging and
broad beam temperatures (\S4).  In $\S5$ we discuss how the tight
correlation between cluster mass and X-ray temperature can be used to
determine binding masses with even greater statistical accuracy.  We
conclude in $\S6$ with a brief discussion of some implications of these
results.

\bigskip
{\bf 2.  Numerical Methods }
\medskip

{2.1  $\underline{ \rm{Sample~Description} }$ }
\smallskip

We use 58 N-body/gas dynamics simulations drawn from three different
projects and run with two completely independent Lagrangian codes. In
all cases we use the Smoothed Particle Hydrodynamics (SPH) technique to
follow the evolution of the gas, and either a $P^3M$ code or a
tree-based N-body code to compute the gravitational interaction between
particles. All the simulations assume a standard, cold dark matter
(CDM) initial fluctuation spectrum with $\Gamma \!\equiv\! \Omega h \se
0.5$.  We neglect the radiative cooling of the gas, as well as magnetic
fields as a possible source of pressure support. We take the baryon
density parameter to be $0.1$, and use a Hubble constant of $50$ km
s$^{-1}$ Mpc$^{-1}$ ($h=0.5$) when scaling to physical units.  Details
of the models can be found in the references quoted below.

The first set consists of 28 runs from Chris Metzler's thesis (Metzler
1995), which examines the effects of energy feedback and mass ejecta
from early type galaxies on the evolution of the intracluster medium
(ICM).  The set consists of 14 clusters of different mass obtained by
constraining the initial density field in cubic, periodic regions
ranging from $25$ to $60$ Mpc (Bertschinger 1987). Each realization is
run twice, once including the effects of galaxy feedback and a second,
control run without feedback.  We shall refer to each of these samples
as ``EJ'' and ``2F'', respectively. All of the runs assume $\Omega=1$.
Each simulation uses a total of $65,536$ particles, divided equally
between gas and dark matter. A description of the feedback
implementation and application to the formation of a Coma--sized
cluster can be found in Metzler \& Evrard (1994). The full set of runs
is described in Metzler \& Evrard (1995).  A salient feature of the
ejection runs is that they employ a rather extreme model of galactic
feedback, in which early-type galaxies lose half their initial mass by
winds.  The EJ and 2F series are intended to define an envelope within
which realistic models of feedback should lie.

The second set of runs is a sample of 24 used to investigate the \xray
cluster morphology--cosmology connection by Evrard \etal (1993) and
Mohr \etal (1995).  For this project, eight different realizations of
the initial density field were evolved within three background
cosmologies in periodic cubes ranging from 30 to 60 Mpc in length.
Again, $65,536$ particles per run were used.  The cosmological models
explored were the standard CDM scenario (model ``EdS''; $\sigma_8 \se
0.59$, $\Omega = 1$), an unbiased, open CDM universe (model ``Op2'';
$\sigma_8 \se 1.0$, $\Omega_0 \se 0.2$, $\lambda_o \se 0$) and an
unbiased, low density universe with a flat geometry (model ``Fl2'';
$\sigma_8 \se 1.0$, $\Omega_0 \se 0.2$, $\lambda_o \se 0.8$).  Here
$\sigma_8$ is the rms mass fluctuation in spheres of $8h^{-1}$ Mpc,
and $\Omega_0$ and $\lambda_0$ are the present values of the density
parameter and the cosmological constant, respectively.  An important
aspect of this set is that the runs corresponding to each cosmology
share common dynamical histories through the use of the same eight
initial density fields.  The benefit of this procedure is that
relative differences between the final characteristics of clusters can
be ascribed to the effect of the different cosmological backgrounds
rather than to ``cosmic scatter'' in the initial density fields.

The final set consists of six models taken from Navarro, Frenk \&
White (1995a; model ``NFW'') which were used to examine the evolution
of scaling laws relating the dynamical and X-ray properties of
clusters of different mass.  These simulations were evolved with a
code completely independent from the one used in the runs described
above; a Tree/SPH code described in detail in Navarro \& White
(1993, 1994). The six clusters were identified at $z=0$ ($\sigma_8=0.63$) in
the cosmological N-body simulations of Frenk \etal (1990), and then
resimulated individually at higher resolution. Each simulation has
$21,296$ particles, half gas and half dark matter, distributed over
boxes of size 15-50 Mpc, depending on the mass of the cluster. The
tidal field due to material surrounding each cluster out to $360$ Mpc
is treated self-consistently by coarse sampling the surrounding matter
with particles of radially increasing mass.

The ensemble of simulations span a wide range in mass and temperature,
from $\sim 10^{14}$ to $3 \times 10^{15} M_{\odot}$, and from $\sim 1$
to $10$ keV.  Table 1 provides a summary for future reference of the
runs and the notation that we use.  The clusters produced in the
different projects have similar spatial and mass resolution
properties. The ratio between the size of a simulated cluster (as
measured by the radius, $\rfiveh$, where the mean density relative to
the critical value is $500$) and the gravitational softening is in the
range $10 \lta \rfiveh/\varepsilon \lta 30$.  The runs have similar
numbers of particles within $r_{500}$, typically 5000 in each
component.

\medskip
2.2  $\underline{\rm Hydrostatic, Isothermal \, Mass \, Estimates}$
\smallskip

The typically smooth morphology of the X-ray emission from the hot,
intracluster medium leads naturally to the hypothesis that the gas is
near equilibrium, stratified along isopotential surfaces in a mildly
evolving distribution of dark matter, gas and galaxies.  The
assumption of hydrostatic equilibrium --- the balance between pressure
gradients and gravity --- for gas supported solely by thermal
pressure, results in a direct measure of the binding mass
$M(r)$. Assuming spherical symmetry,
$$
M(r) \ = \ - {kT(r) \over G \mu m_p} \ r \
\biggl( {\dlog \rho(r) \over \dlog r} + {\dlog T(r) \over \dlog r} \biggr)
\eqno(\ino)
$$
where $\rho(r)$ and $T(r)$ are the gas density and temperature
profiles, $k$ is Boltzmann's constant, and $\mu m_p$ is the mean
molecular weight of the gas. In principle, all terms on the right hand
side of this equation are measurable.  The main limitation is that one
must deconvolve three dimensional profiles from two dimensional
surface brightness information.  This requires knowledge of, or a
model for, the temperature profile $T(r)$.  Since direct measurements
of the temperature as a function of radius $T(r)$ (more precisely, the
X-ray emission-weighted projected temperature) are still a relatively
rare commodity, the common practice has been to assume that the gas is
isothermal at the spatially averaged temperature $T_{\rm X}$
determined from a broad beam spectroscopic instrument such as {\sl
EXOSAT} or {\sl Ginga}.  Preliminary ASCA results (Ikebe \etal 1994;
Mushotzky 1994), as well as previous direct measurements (see, e.g.,
Watt \etal 1992) and the numerical models that we use here generally
support this assumption (see $\S3.3$ below).

The usual parametrization of the density profile of the ICM is based
on the isothermal, $\beta$-model proposed by Cavaliere \&
Fusco-Femiano (1976), $\rho_{\rm ICM}(r) \se \rho_o (1+(r/r_c)^2)^{-3
\beta/2}$. With this assumption, equation (1) reduces to
$$
M(r)= \ {3 \beta \over G} \ {kT_{\rm X} r\over \mu m_p} \ {(r/r_c)^2
\over 1+(r/r_c)^2} = 1.13 \times 10^{15} \beta \, {T_{\rm X} \over 10 \kev}
\ {r \over \mpc} \ {(r/r_c)^2 \over 1+(r/r_c)^2} \, \msol ,
\eqno(\ino)
$$
assuming $\mu \se 0.59$ for the second equality.
For an isothermal gas, the \xray surface brightness as a function of
projected radius $S_X(\theta)$ simply
reflects the integrated emission measure and can be expressed as
$$
S_X(\theta) \ = \ S_0 \ (1+(\theta/\theta_c)^2)^{-3\beta+1/2} .
\eqno(\ino)
$$
X-ray spectroscopy provides $T_{\rm X}$, while estimates of $\beta$ and $r_c$
are obtained from \xray imaging.

To mimic the mass estimates derived from X-ray observations, we
generate for each simulated cluster three artificial ROSAT images and
emission weighted temperatures $T_{\rm X}$ along the principal axes of
the volume.  Generating synthetic ROSAT images requires a choice of
several parameters, including: (i) the cluster redshift $z$, (ii) the
exposure time $t_{exp}$, and (iii) the background noise level $S_N$.
Fitting the resulting surface brightness profiles to the $\beta$--model
introduces two additional parameters; (iv) the background subtraction
level $S_b$ and (v) the minimum flux level to which the fit is
performed $S_{min}$.  An additional choice is the energy band of the
observations, which we take to be the full ROSAT band $0.1$-$2.4
\kev$. Because there is no obscuring galaxy in our `observations', our
results are insensitive to the exact choice of the lower energy limit
for the band employed. We use emission weighted temperatures within
circular apertures large enough to contain nearly all the cluster
emission.  Since most of the photons come from the central regions of
the cluster, the estimates of $T_{\rm X}$ are rather insensitive to
the aperture choice.

We use a single parameter set to construct the synthetic images of all
model clusters, chosen to be representative of typical \xray
observations.  We view the clusters at a redshift $z \se 0.04$, image
them for $t_{exp} \se 7200$ s, include a Poisson background in the
counts at a level of $3 \times 10^{-4} \cntflux$, subtract a mean
background with exactly this value, and fit each to the $\beta$--model out
to the radius at which the corrected counts reach the background
level.  We use emission weighted temperatures in the same ROSAT band,
measured within an angular scale of $16^\prime$, corresponding to a
fixed metric radius of $1.04 \mpc$.  These
parameter values are summarized in Table 2.  We investigated
alternative parameter sets, and found that the results depended mainly
on the image quality through the combination of redshift, exposure
time and assumed background.  The parameter choices above provide high
signal-to-noise images for the majority of the sample.

For each cluster, the three images and emission weighted temperature
maps are generated and reduced to obtain the values of $\beta$, $r_c$
and $T_{\rm X}$. This procedure works very well for clusters formed in
cosmological models with $\Omega=1$.  However, as discussed in Mohr
\etal (1995), in low density universes the emission profiles of the
cluster models are strongly peaked, resulting in unacceptably poor
$\beta$--model fits. In this case, we decided to excise the central
regions and to fit only data outside a region of angular radius
$4\arcmin$, corresponding to a linear scale of $260$ kpc (see Mohr
\etal 1995 for details).  Typical statistical uncertainties in the
fitted parameters are $\lta 5\%$, but in a few cases the uncertainties
can be as high as $\sims 40 \%$. Once $\beta$, $r_c$ and $T_{\rm X}$
have been determined, we can estimate the binding mass as a function
of radius using equation (2).
%{\bf Gus: Is this ``excise'' done also
%when computing $T_X$ for the open model runs?}

\medskip
{2.3  $\underline{\rm {A~Particular~Example}}$ }
\smallskip

Before examining results for the entire sample, we discuss data for a
particular cluster which highlight characteristics typical of the
ensemble. Figure~1 shows \xray surface brightness and emission
weighted temperature maps, as well as the emission weighted
projected velocity field for
the ICM in a cluster taken from the EdS sample.  The field of view in
each of the panels is $64^\prime$, corresponding to $4.2 \mpc$ at the
assumed cluster redshift of $0.04$.  Two of the three projections show
a bimodal central structure; the result of a recent merger involving
two main sublumps with mass ratio $2.8$:$1$.  Infall patterns of the
sublumps are evident in the velocity field of the middle ($y$-axis)
projection.  The temperature map shows significant variations (up to a
factor $\ssim 2$) in temperature on scales of a few hundred kpc.  The
hot spots occur in regions where the gas is being compressed and
mildly shocked by the interpenetrating subcluster cores.  Cooler gas
can be seen trailing in the wake of the substructure cores.

Several of the features of this simulated cluster, particularly the
$y$-axis view, are reminiscent of the cluster A2256.  Features in
common include a bimodal central structure and spatial variations in
temperature similar in morphology and amplitude (Briel \etal 1991;
Miyaji \etal 1993; Briel \& Henry 1994).  Furthermore, an offset,
extended radio halo exists in A2256 which strongly indicates the
presence of a flow pattern similar to that seen in the $y$-axis
image (Rottgering \etal 1994).

Figure~2 shows the azimuthally averaged \xray surface brightness and
emission weighted temperature profiles for the three projections.
Surface brightness profiles are scaled down successively by an order
of magnitude for the sake of clarity; as are the temperature
profiles by factors of 2.  Values of the fitted parameters are $r_c
\se 294 \pm 6$, $384\pm 10$, and $359\pm9$ kpc; and $\beta \se 0.89\pm
0.02$, $0.85 \pm 0.02$, and $0.81 \pm 0.02$ for the three projections
from top to bottom, respectively.

Although significant spatial variations in temperature are obvious in
Figure~1, the radially averaged temperature profile is close to
isothermal over a significant fraction of the cluster image.  The
temperature varies by $\lta 20\%$ within $10^\prime$, a region wherein
the surface brightness drops by a factor of 30.  The temperature drops
by a factor of 2 at about $20^\prime$ from the center, but the surface
brightness at this radius is already below the adopted background of
$3 \times 10^{-4} \cntflux$.

{}From Figure~1 we learn the
importance of examining spatial temperature maps directly, since a
flat azimuthally averaged profile need not be indicative of a truly
isothermal ICM.  Comparison of the temperature and surface brightness
maps can provide useful dynamical clues, although geometry plays an
obscuring role.  From the single dynamical configuration corresponding
to the cluster in Figure~1 one can get, depending on projection (from
top to bottom), (i) a fairly relaxed \xray image with asymmetric
temperature map and strong velocity gradients; (ii) a bimodal \xray
image with an ``S''--shaped hot spot resulting from a symmetric infall
pattern and (iii) a bimodal \xray image with a peanut--shaped hot spot
and a relatively modest velocity field in projection.

\bigskip
{\bf 3.  Is the ICM Hydrostatic and Isothermal ? }
\medskip

The assumption of hydrostatic equilibrium underlies the \xray mass
estimate method, and the assumption of isothermality greatly
simplifies it.  In this section, we analyze how close to hydrostatic
equilibrium and isothermality the cluster models are by inspecting
directly their three dimensional velocity and temperature fields.

\medskip
3.1 $\underline{ \rm The~ICM~Velocity~Field }$
\smallskip

Figure~3 shows the radial Mach numbers $\langle v_r \rangle/c_s$ and
$\langle v_r^2 \rangle/c_s^2$ derived from the gas sound speed
$c_s^2(r) \se 5 kT(r)/3\mu m_p$ and the first and second moments of
the radial velocity field, respectively.  To facilitate comparison
between clusters of different sizes, the mean interior density
contrast $\delta_c(r) \se 3 M(r)/4\pi \rhocrit r^3$ is used as the
radial variable in the figure, reversed to reflect the
correspondence with cluster radius.  Note that the density contrast
used here is defined with respect to the critical value for closure
$\rhocrit=3\Ho^2/8 \pi G$ in all the models.
The center of the cluster is defined
as the position of the most bound dark matter particle, and velocities
are calculated with respect to the mean velocity of all cluster
particles linked by a standard friends--of--friends algorithm using a
linking parameter $0.15$ times the mean interparticle spacing.  The
$\Omega \se 1$ runs all exhibit a common structure, with the exception
of two of the NFW runs and one EdS run, which are undergoing strong
mergers at $z=0$. The Mach numbers of these systems are significantly
higher than the average.  The low $\Omega$ runs have generally
quieter velocity fields, as expected given their earlier formation
times and, consequently, their dynamical maturity relative to clusters
formed in a high density universe.

The mean radial Mach number $\langle v_r \rangle /c_s$ has the
characteristic signatures expected from gravitational infall (Gunn \&
Gott 1972).  An outer zone of mildly supersonic infall surrounds the
`virial' region of the cluster, within which the gas has been largely
thermalized and is close to hydrostatic equilibrium.  The infall
regime is largely absent in the low $\Omega$ models, due to the
stagnated growth of linear perturbations on large scales.  This,
however, does not imply that there are no recent merger events in the
open universe sample --- at least one Op2 cluster is experiencing an
ongoing merger event at the present time.

The right--hand panels of Figure~3 provide an upper limit on the ratio
of kinetic to thermal pressures for the gas.  This ratio rises
monotonically with radius, from values $< \! 10\%$ for radii where
$\delta_c > 500$ to values $\gta 50\%$ at radii where $\delta_c \lta 100$.
A few ongoing mergers are clearly recognized by the large values of
the $\langle v_r^2 \rangle/c_s^2$ ratio near the center. In this case,
interpreting the ratio between kinetic and thermal energy as the
relative contribution of kinetic and thermal pressures to support the
gas does not apply, since the systems are far from equilibrium.  As
the velocity field in Figure~1 indicates, non--zero values of $\langle
v_r^2 \rangle$ arise during mergers from large--scale bulk motions of
the gas across the face of a given radial shell rather than from a
local, uniform dispersion on the shell.

{}From Figure~3, we derive a value $\delta_c \se 500$ as a conservative
estimate of the boundary between the inner, virialized region of the
clusters and their recently accreted, still settling outer envelopes.
Define $r_{\delta_c}$ as the radius within which the mean interior
density is $\delta_c$ times the critical value.  Then, within $\rfiveh$,
the hydrostatic equilibrium assumption is valid since the gas is, on
average, neither expanding nor contracting.  This estimate is
conservative in the sense that, in many clusters, hydrostatic balance
appears to hold even at somewhat larger radii.  The turnaround radius
can be seen to occur at radii about a factor $3-4$ larger than
$\rfiveh$ (for $\Omega=1$).  This is consistent with the spherical
infall models of Bertschinger (1985) in which infalling gas is shocked
nearly to rest at a radius about a third of the turnaround radius.
Despite this nice agreement, we stress that the accretion in these
three dimensional models is far from spherically symmetric.

Table 3 gives mass averaged values of the two radial Mach numbers
measured within $\rfiveh$ for each set of runs.  The mean radial Mach
numbers are all quite small, typically a few percent or less, and are
consistent with zero given the measured error in the mean.  Again, the
NFW set seems to be the most dynamically active, as can be seen from
the measures of both velocity moments.  Typical values of $\langle
v_r^2 \rangle/c_s^2$ are $ \lta \! 10\%$, indicating that the gas is
hydrostatically supported by thermal pressure to this accuracy within
$\rfiveh$.

\medskip
\vfill\eject
3.2 $\underline{ \rm Cluster~Scaling }$
\smallskip

As discussed above, we will use the radius $\rfiveh$ as a
characteristic length scale separating the nearly hydrostatic central
region of a cluster from the surrounding, recently accreted outer
envelope.  As shown by Navarro \etal (1995a,b) and Metzler \& Evrard
(1995), clusters of different
mass have similar structures when scaled to such a characteristic
radius. This similarity, along with the equilibrium assumption
validated above, implies a power--law relationship between cluster
`size' and temperature
$$
T \ \propto \ {M(<\rfiveh) \over \rfiveh} \ \propto \ \rfiveh^2,
\eqno(\ino)
$$
as appropriate for systems of similar density in virial
equilibrium.  Figure~4 shows this relationship for the simulations,
using the measured value of $\rfiveh$ and three orthogonal measures of
the emission weighted temperature $T_{\rm X}$.  Each cluster appears
three times in this plot; the dispersion in $T_{\rm X}$ for different
projections is typically quite small. Fitting the results with a power
law, we find
$$
\rfivest(T_{\rm X}) \ = \ 2.48 \pm 0.17 \
\biggl({T_{\rm X} \over 10 \kev}\biggr)^{1/2} \mpc
\eqno(\ino)
$$
where the quoted uncertainty in the intercept is given by the standard
deviation of the residuals in the log space fit.   The actual best fit
slope for each individual set of runs differs by less than $10\%$ from
the $0.5$ exponent expected from equation (4).  There are small offsets
between the models.  The ejection models, for example, have $\sims 6\%$
smaller $\rfiveh$ at a given temperature or, equivalently, $\sims 12\%$
higher emission weighted temperatures for a given $\rfiveh$, compared
to the 2F runs.  The slope for the EJ sample is also slightly ($\sims
10 \%$) steeper than those of the non-ejection samples, as expected
from the differential effect of feedback, which raises the temperature
of poor cluster gas proportionally more than that of rich clusters
(Metzler \& Evrard 1995).  There is no significant difference between
the NFW, 2F and EdS data sets, indicating an encouraging agreement
between the results of models run with completely independent codes for
the same cosmological model.

\medskip
3.3 $\underline{ \rm Temperature~Profiles }$
\smallskip

Figure 5 shows the mass-weighted gas temperature, in units of $T_{\rm
X}$, as a function of the normalized radius $r/\rfiveh$ for all runs.
For $\Omega=1$, the profiles are close to isothermal within
$\rfiveh/2$, and decline gently beyond that radius; the temperature at
$\rfiveh$ is $\sims 20\%$ lower than at the center. In all cases, the
modest drop in temperature within $r_{500}$ is due to the fact that the
density profiles of both gas and dark matter are slightly steeper than
isothermal in their outer parts (Navarro \etal 1995a,b).  In the case
of an open universe, the profiles are noticeably steeper; the
temperature at $\rfiveh$ is on average a factor of two lower than the
central value. This comes as no surprise, for the density profiles of
clusters formed in low-density universes are expected to be
significantly steeper than those formed in an $\Omega=1$ cosmology
(Hoffman 1988, Crone, Evrard \& Richstone 1994).  Since \xray data
rarely extend significantly beyond $\rfiveh$, an isothermal assumption
should be appropriate for most observed clusters.

\bigskip
{\bf 4. Mass Estimates from the $\beta$-model.}
\medskip

After constructing the synthetic ROSAT images as described in $\S2.2$,
fitting the surface brightness with equation (3), and measuring the
projected X-ray emission weighted temperature, we derive estimates of
the cluster binding mass as a function of radius via equation (2).
Once again, in order to compare clusters of different mass (and size),
we transform the radial variable into an {\it estimated} density
contrast $\delta_c^{est}$ relative to the critical density
using the estimated mass $M^{est}(r)$ from eq.(2);
$$
\delta_c^{est}(r) \ \equiv  \
{ {3 M^{est}(r)} \over {4 \pi \rhocrit r^3} } \ = \
{ 2 G M^{est}(r) \over {\rm H}_0^2 r^3 } \ = \
  \biggl({2 \sigma(r) \over {\rm H}_0 r}\biggr)^{2}
\eqno{(\ino)}
$$
where $\sigma^2(r) \se GM^{est}(r)/2\, r$ is roughly the
one--dimensional velocity dispersion of the cluster.  For a rich
cluster with $\sigma \se 1000 \kms$ at $r \se 1 \hinv \mpc$, the
estimated density contrast is $\delta_c^{est} \se 400$.  Note that
$\delta_c^{est}$ depends on the combination $H_0 r$, and therefore it
is independent of the Hubble constant, making it a useful measure of
radius for observations.

Figure~6 shows the ratio between estimated and true mass as a function
of the radial coordinate $\delta_c^{est}$.  The finite dynamic range in
the simulations limits the density contrast to values $\lta 5000$.
This figure shows that the cluster binding masses are on average quite
accurately determined at overdensities between 500 and 2500.  In the
outer regions, where $\delta_c^{est} \lta 200$, masses are typically
overestimated because the estimated mass, assumed to increase linearly
with radius, increases with radius faster than the true mass in this
region. The effect is more pronounced in the low density runs, where
the cluster density profiles are steepest.  Overestimates by factors up
to $3$ are seen in the low-density runs at $\delta_c^{est} \se 100$.

Figure~7 presents histograms of the estimated-to-true mass ratio at
radii corresponding to $\delta_c^{est} \se 2500$, $500$ and $100$.
These three values of the density contrast sample dynamically different
regions and span a range comparable to that of current \xray
observations.  Dashed vertical lines in the figure show $\pm 40\%$
error for reference.  The trend toward overestimates at low density
contrasts noted in Figure~6 is apparent in the rightmost column.  The
omission of the NFW runs at this contrast is technical in origin; these
simulations do not extend reliably to these low overdensities.

Relatively large (factor $\sims 2$) underestimates occur in a few of
the $\Omega \se 1$ clusters.  Three of the worst offenders arise from
images of a single, strongly bimodal EdS cluster which is currently
experiencing a major merger. Indeed, ongoing mergers are responsible
for six of the worst underestimates at $\delta_c^{est}=500$.  Synthetic
ROSAT images of these cases are shown in Figure~8. Bold and light
circles indicate the estimated and true values of $\rfiveh$.  The
complex, multi-peaked structure of the \xray emission in these images
is a strong signal of dynamical unrest.  Although such objects would
probably be rejected by observers attempting to apply an equation based
on hydrostatic equilibrium, we include these cases in our analysis
below since they only represent a small fraction of the total
population.

The similarity between the $\Omega \se 1$ sets suggests that co-adding
the runs is appropriate in order to compute the ensemble statistics.
Figure~9 shows the histogram of estimated-to-true mass ratios, ${\cal
X} \sequiv M^{est}/M^{true}$, evaluated at $\delta_c^{est}=500$ for the 126
images of the combined EJ, 2F, NFW and EdS sets.  The histogram is
nearly Gaussian with mean ${\bar {\cal X}} \se 1.02$ and standard
deviation $\sigma_{\cal X} \se 0.29$.  At this density contrast, the
$\beta$-model estimates are unbiased and have rather modest scatter.
We repeated this procedure at contrasts $\delta_c^{est}
\se 2500$, 1000, 250 and 100.  The means and standard deviations of
the resulting estimated-to-true mass ratio distributions are given in
Table~4 for the $\Omega \se 1$ and $\Omega_0 \se 0.2$ ensembles.
Values in this Table reflect the trends apparent in Figures~6 and 7;
the bias and the variance in the mass estimator both increase with
increasing radius (\ie, towards lower density contrasts).  The increase
in the variance with radius is probably linked to the longer dynamical
times of the outer regions.

Although the estimator does worse in the case of strongly bimodal
clusters (see Figure~8), there are cases where the geometry of the
projection, together with the interplay between the estimated values
of $\beta$, $r_c$, and $T_{\rm X}$, can result in accurate mass
estimates even for clusters with suspicious looking
\xray images.  As an example, consider the cluster shown in Figure~1.
The $\delta_c^{est}=500$ mass estimates for the three orthogonal
projections yield (from top to bottom) ${\cal X} \se 1.17$, $1.02$ and
$0.97$.  The most symmetric \xray image incurs the largest error while
the two images displaying core bimodality are more accurately
determined.  Note that $\rfiveh$ for this cluster is $1.62 \mpc$, or
$24^\prime$ in the figure, well beyond the core region.  The larger
value in the top projection compared to the others is due to slightly
larger values of $\beta$ and $T_{\rm X}$ and a smaller core radius
compared to the other projections.  These values
result from the fact that the line of sight in the top panel is
nearly parallel to the collision axis of the penetrating cores.

In summary, we find $\beta$-model mass estimates to be nearly unbiased
and accurate to a few tens of percent in the regime $250 \lta
\delta_c^{est} \lta 2500$ for the $\Omega \se 1$ models and
$\delta_c^{est} \gta 1000$ for the $\Omega_0 \se 0.2$ sample.  A bias
toward overestimating masses exists at low values of $\delta_c^{est}$.
Clusters with strongly bimodal or more complex images involve the
largest mass underestimates.  Because of the interplay of $T_x$,
$\beta$ and $r_c$, there is not a simple, general connection between
the properties of the \xray image and the accuracy of the mass
estimates obtained with the $\beta$-model.

% 0.0683428 1.73638  error, rdelta for EdS run  c40b  1   -> Mest/Mtrue=1.17
% 0.0101851 1.63762  error, rdelta for EdS run  c40b  2          1.02
% -0.0137625 1.59861  error, rdelta for EdS run  c40b  3	 0.97

\bigskip
\vfill\eject
{\bf 5. Estimates Based on Cluster Scaling Relations.}
\medskip

As discussed in $\S3.2$, massive clusters within a given cosmology
exhibit a remarkably similar structure when scaled to a fixed density
contrast.  Together with the
condition of virial or hydrostatic equilibrium, this implies that the
temperature of a cluster is, on its own, a good indicator of the size
and mass of the system. Figure~4 shows this result very clearly. The
tight correlation shown in this figure between $T_{\rm X}$ and
$r_{500}$ implies a similarly tight correlation between mass and
temperature since, by definition, the mean density within $r_{500}$ is
500 times the critical density. Armed only with the broad beam
temperature measure $T_{\rm X}$, we can thus form an estimate
$$
M_{500}^{est}(T_{\rm X}) \ \equiv \ 500 {4 \pi \over 3} \rhocrit
({\rfivest})^3 \ = \ 2.22 \times
10^{15} \, \biggl({T_{\rm X} \over 10 \, {\rm keV}}\biggr)^{3/2}
M_{\odot}
\eqno{(\ino)}
$$
for the mass within the radius $\rfivest(T_{\rm X})$ given by equation
(5).

We compare this mass estimate with the true mass within
$\rfivest(T_{\rm X})$ for
the $\Omega \se 1$ ensemble in Figure 10.  (Results for the $\Omega_0
\se 0.2$ sets are similar.)  The distribution of estimated-to-true
mass ratios is nearly Gaussian with a standard deviation of
only $15\%$.  Note that, in this case, no cluster in the $\Omega=1$
sets has its mass over or underestimated by more than $40 \%$,
regardless of its dynamical state.  This procedure can be extended to
other values of the density contrast in a straightforward way.  For a
given $\delta_c$, we compute the characteristic radii $r_{\delta_c}$ from
the numerical sample and fit them to a relation of the form
$$
r_{\delta_c}(T_{\rm X}) \ = \ r_{10}(\delta_c) \
\biggl({T_{\rm X} \over 10 \kev}\biggr)^{1/2}
\eqno(\ino)
$$
where the normalization $r_{10}(\delta_c)$ is the average radial scale of
10 keV clusters at density contrast $\delta_c$.

The resulting distributions of ${\cal X}$ are unbiased by
construction and have standard deviation  $\sigma_{\cal X}$.
Table~5 shows the characteristic radii $r_{10}(\delta_c)$ and the scatter
in the mass estimator $\sigma_{\cal X}$ for density contrasts
$\delta_c=$100, 250, 500, 1000 and 2500.  Slight offsets in the
characteristic radii are evident between the high and low $\Omega$
cosmologies, consistent with the difference in cluster density
profiles.  As in the $\beta$-model estimates, the scatter in the mass
estimates increases with radius.  At density contrasts of a few
thousand, the uncertainty in the mass estimates is extremely small
$\sigma_{\cal X} \lta 10\%$.

%We note here a slightly technical point explaining the superscript
%on $\rfiveh$ in equation (7).  Given $T_{\rm X}$, we consider the
%radius $\rfivest$ as the estimated value of $\rfiveh$ within
%which one forms the mass estimate via equation (9).  For the numerical
%models, however, we have direct knowledge of the true value of
%$\rfiveh$ and the mass within that radius $M(\rfiveh)$.  Figure~10
%compares $M_{500}^{est}$ to the true mass within the estimated radius
%$M(\rfivest)$ rather than to $M(\rfiveh)$, the true mass
%within the actual value of $\rfiveh$.  For the latter, one obtains a
%distribution with a somewhat larger variance than the
%former.

Another way to interpret these results is to consider that the scaling
law mass estimate is consistent with the hydrostatic, $\beta$-model
estimate, equation (2), when evaluated at $\rfivest(T_{\rm X})$ with a
characteristic value of $\beta$ given by
$$
\beta_\ast \ = \ 0.79 \ (1+(r_c/r)^2) .
\eqno(\ino)
$$
The second term is a typically small ($\lta 5\%$) correction at
$\delta_c \se 500$.  This value for $\beta$ compares well with
measured values of $\beta$ for many well studied, rich \xray
clusters, among them Coma (Hughes 1989) and
A2256 (Briel \etal 1992).

Why are the scaling law estimates more accurate than those of the
$\beta$-model?  Consider the variance in the $\beta$-model mass
estimated at a fixed radius
$$
\biggl( {\Delta M \over M} \biggr)^2 \ = \
\biggl( {\Delta T_{\rm X} \over T_{\rm X}} \biggr)^2 + \
\biggl( {\Delta \beta \over \beta} \biggr)^2
+ \ 2 \biggl( {\Delta T_{\rm X} \over T_{\rm X}} {\Delta \beta \over \beta}
\biggr) .
\eqno(\ino)
$$
In a universe filled with perfectly hydrostatic, self-similar
clusters, all clusters would follow a $r_{\delta_c}(T_{\rm X})$ relation like
equation (8) exactly and all would have a fixed value
$\beta_\ast$ for their outer profile slope.
Introduction of perturbations in temperature and
density off this perfect sequence will lead to a non--zero variance in
the mass estimate.  The only way to retain perfect mass estimates is
to introduce correlated perturbations $\Delta \beta/\beta \sequiv
- \Delta T_{\rm X}/T_{\rm X}$ in density and temperature, so that hydrostatic
equilibrium is maintained.  Uncorrelated perturbations in
$\beta$ and $T_{\rm X}$ will lead to a larger variance than that arising
from perturbations in one parameter alone.

Figure~11 shows the perturbations measured directly in the
simulations.  Perturbations in $\beta$ are defined with respect to the
average, $\beta_\ast =0.79$, while those in temperature are
defined with respect to the mean radius--temperature  relation, using
the known value of $\rfiveh$ to define the unperturbed cluster
temperature for each cluster.  It is clear from this Figure that the
data exhibit no correlation between $\Delta \beta/\beta$ and $\Delta
T_{\rm X}/T_{\rm X}$.  The larger variance in the $\beta$-model
compared to the scaling law mass estimates can thus be understood as
arising from an additional, independent source of error in the
$\beta$-model estimator; {\sl the introduction of $\beta$ from \xray
imaging is essentially adding noise to the mass estimates}.  As a
concrete example, consider again the three images in Figure~1.  The
$\beta$-model mass estimates yield ${\cal X} \se 1.17$, $1.02$ and
$0.97$ from top to bottom whereas the scaling law results are ${\cal
X} \se 1.05$, $1.00$ and $0.99$, respectively.

The ``noise'' added by the measured values of $\beta$ has several
sources.  Recent dynamics play a role.  Clusters are not in exact
hydrostatic balance, particularly in their outer parts.  The present
density and temperature structure can be perturbed by prior mergers
and accretion.  Geometry also plays a role.  The cluster gas
distribution is, in general, ellipsoidal rather than spherical
and the measured values of
$\beta$ are obtained from projected, two-dimensional images of the
three-dimensional density distribution.  Finally, values of
$\beta$ derived from surface brightness fits are sensitive to
contamination from foreground/background sources, choice of cluster
center, as well as image quality and technical aspects of data
reduction procedures.  These concerns are compounded when one
considers the basic fact that $\beta$ is a measure of the {\it
derivative} of the logarithm of the surface brightness.

The scaling law method is superior to the $\beta$-model because of its
smaller variance.  Its accuracy is also remarkably insensitive to the
dynamical state of clusters and the cosmological background.  Its main
drawback is the reliance on numerical experiments to provide the
normalization, $r_{10}(\delta_c)$, of equation (8) which, in turn,
depends on the particular structure formation model under scrutiny, and
may depend as well on the numerical method used in the investigation.
Regarding the latter, we are encouraged by the good agreement shown
between two independent codes used in this study.  We anticipate future
studies by groups employing independent numerical methods will address
the robustness of the normalization $r_{10}(\delta_c)$.  We also find
room for optimism in the rather modest sensitivity to $\Omega_0$
displayed by the normalization and scatter in mass estimates shown in
Table~5.  This insensitivity may be rooted in the fact that the scaling
laws merely reflect the condition of virial equilibrium within
$\delta_c \ssim {\cal O}(10^3)$.  Overall, these results show that it
is possible to use \xray spectroscopy to estimate cluster masses with
an {\it rms} accuracy substantially better than $20\%$.

\bigskip

% \vfill\eject

{\bf 6. Discussion and Conclusions.}
\smallskip

The results described in the previous sections agree well with those of
Schindler (1995), who found biases and variance of a few tens of
percent or less in the ``normal cluster'' sample derived from the
numerical simulations of Schindler \& B\"ohringer (1993) and Schindler
\& M\"uller (1993).  The simulations and analysis methods in those
works differ in many respects from those used here. In particular, the
Eulerian gas dynamics scheme adopted in their study captures shocks
more accurately than the SPH technique used here, while the adaptive
nature of the SPH smoothing kernel provides better spatial and mass
resolution in high density regions.  Given the significant differences
between the independent simulation algorithms used in these studies, we
find the degree of qualitative and quantitative agreement rather
encouraging.

Tsai, Katz \& Bertschinger (1994) also examined the accuracy of the
$\beta$-model mass estimates applied to an SPH simulation which
included radiative cooling for the gas, but which excluded the effects
of galaxy formation and feedback. At 1 Mpc from their cluster center,
where $\delta_c \simeq 700$, they found that the $\beta$-model
overestimated the mass by $\sims 25\%$, typical of the uncertainties
found in our analysis.  They also found large underestimates at radii
near the core radii of their surface brightness fits ($\sims 200 \kpc$)
which arose, in part, from the presence of a strong gradient in the gas
temperature near the center. However, at these high density contrasts,
their results are compromised by the artificially strong concentration
of baryonic material near the center.

The improved variance of the scaling law estimates over the
$\beta$-model is a mixed blessing.  A serious concern is the
sensitivity of the slope and normalization of the $r_{\delta_c}-T_{\rm
X}$ relation to the assumed cluster physics and the underlying
cosmological model.  We find very modest sensitivity to $\Omega_0$ near
$\delta_c \ssimeq 10^3$ (Table~5) and find that the ejection models do
not differ substantially from the infall sample.  We suspect, and
this suspicion is supported by collisionless simulations of cluster
formation (Crone \etal 1994), that the impact of changing the
initial perturbation
spectrum will be less than that of varying $\Omega_0$, but this issue
remains to be investigated in detail.

What should then an observer with \xray data do to estimate a cluster's
mass?  It depends on whether it is more important to minimize the bias
or the variance in the mass estimate.  For example, when comparing a
set of \xray derived masses to values derived using an independent
method (\eg, weak gravitational lensing), minimizing the bias in the
\xray binding masses would perhaps be most important, and the
$\beta$-model approach preferred.  On the other hand, if one were
looking for the slope of correlations between an observable cluster
property (\eg, optical luminosity or velocity dispersion) and binding
mass, then minimizing the variance would be more important, and the
scaling law method preferred.

To summarize, we have used an ensemble of 58 gas dynamics cluster
simulations to investigate the accuracy of binding mass estimates based
on the hydrostatic, isothermal $\beta$-model and on the
temperature--mass relation.  We have also analysed the velocity and
temperature fields of the numerical models to address the questions of
hydrostatic equilibrium and isothermality.  A summary of our main
results follows.

\smallskip

\item{$\bullet$} {
Within a radius, $r_{500}$, where the cluster mean interior density is
$500\rhocrit$, the gas velocity field is extremely quiet
(Table~3 and Figure 3), validating the basic assumption of
hydrostatic support by thermal pressure.  Despite local variations in
temperature due to ongoing merger events (Figure~1), the radially
averaged gas temperature is nearly isothermal within $r_{500}$ in the
$\Omega \se 1$ sample.  The $\Omega_0 \se 0.2$ clusters exhibit a
moderate, negative radial temperature gradient.  }

\item{$\bullet$} {
The standard $\beta$-model mass estimator (eq.~4) is nearly unbiased
and has a modest scatter in regions where the mean estimated density
contrast is in the range $500 \lta \delta_c^{est} \lta 2500$.  For
example, at $\delta_c^{est} \se 1000$, the mean value of $M^{est}/M^{true}$
is $0.94$ ($1.15$) with standard deviation $0.23$ ($0.19$) for the
$\Omega \se 1$ ($\Omega_0 \se 0.2$) sample.  The bias and scatter both
increase with cluster radius (decreasing $\delta_c$).  The bias
increases because the true density profiles are steeper than the
assumed isothermal value, while the dispersion increases because of the
longer dynamical timescales characteristic of larger radii. }

\item{$\bullet$} {
We find a strong correlation between $r_{\delta_c}$, the radius
encompassing a mean density contrast $\delta_c$, and $T_{\rm X}$, the
broad beam, emission-weighted gas temperature.  This scaling --- a
reflection of the similarity between clusters of different mass and
their near virial equilibrium state --- can
be used to generate mass estimates with smaller variance than that of
the $\beta$-model.  The degree of scatter is surprisingly small.  At
$\delta_c \se 1000$, the standard deviation is $0.11$ ($0.12$) for the
$\Omega \se 1$ ($\Omega_0 \se 0.2$) sample.  The larger dispersion in
the $\beta$-model method arises because two parameters, $\beta$ and
$T_{\rm X}$, contribute independent sources of error whereas the
scaling law method incurs error from only one parameter, $T_{\rm
X}$. }

\smallskip

The results from the experiments reported here and those from other
experiments cited above practically rule out the possibility of large
systematic errors in the mass determination of galaxy clusters. The
large baryon fractions measured in clusters therefore remain difficult
to reconcile with standard primordial nucleosynthesis in an $\Omega=1$
universe.

\bigskip\bigskip
This research was supported by NASA through Grants NAGW-2367 and
NAG5-2790, by a NATO International Travel Grant and by the NSF through
Grant No. PHY94-07194.  AEE and JFN are grateful for the hospitality
provided by the ITP at UCSB during the ``Radiation Backgrounds and
Galaxy Formation'' workshop and thank D. Bond, C. Norman, J. Ostriker
and S. White for their organizational efforts.  C.A.M. acknowledges
support from a Rackham Predoctoral Fellowship
and a Sigma Xi Grant-In-Aid of Research at the University of Michigan, and
the Department of Energy and NASA Grant NAG5-2788 at Fermilab.

\vfill\eject

\centerline{\bf Table 1}
\smallskip
\centerline { Summary of Model Notation }
\medskip
\vbox{\hbox to \hsize{\hfil\vbox{\halign {#\hfil&&\quad\hfil#\hfil\cr
\noalign{\hrule}\cr
\noalign{\smallskip}
\noalign{\hrule}\cr
\noalign{\medskip}
Label & \hfil $N$ & Description \hfill & Reference \hfill \cr
\noalign{\smallskip}
\noalign{\hrule}\cr
\noalign{\medskip}
EJ & \hfil 14 & $\Omega \se 1$, $\sigma_8=0.59$, wind ejection included
\hfill & Metzler \& Evrard (1994,5) \hfill \cr
2F & \hfil 14 & same as EJ without winds \hfill
& \hfil \H \hfil \cr
NFW & \hfil 6 & $\Omega \se 1$, $\sigma_8=0.63$, Tree/SPH code
 \hfill & Navarro \etal (1995a) \hfill  \cr
EdS & \hfil 8 & $\Omega \se 1$, $\sigma_8=0.59$, no feedback \hfill &
Evrard \etal (1993) \hfill \cr
Fl2 & \hfil 8 & $\Omega_0 \se 0.2$, $\lambda_o \se 0.8$,
$\sigma_8=1$, same IC's as EdS \hfill & \hfil \H \hfil \cr
Op2 & \hfil 8 & $\Omega_0 \se 0.2$, $\lambda_o \se 0$,
$\sigma_8=1$, ~~~~~~~~~ \H \hfill & \hfil \H \hfil \cr
\noalign{\medskip}
\noalign{\hrule}\cr
\noalign{\medskip}}}\hfil}}

\bigskip

\centerline{\bf Table 2}
\smallskip
\centerline { Synthetic Observation Parameters }
\medskip
\vbox{\hbox to \hsize{\hfil\vbox{\halign {#\hfil&&\quad\hfil#\hfil\cr
\noalign{\hrule}\cr
\noalign{\smallskip}
\noalign{\hrule}\cr
\noalign{\medskip}
Parameter & \hfil value \cr
\noalign{\smallskip}
\noalign{\hrule}\cr
\noalign{\medskip}
source redshift $z$ & \hfil 0.04 \cr
exposure time $t_{exp}$ (sec) & \hfil 7200 \cr
background noise level ${S_N}^a$ & \hfil $3 \times 10^{-4}$ \cr
background subtraction level ${S_b}^a$ & \hfil $3 \times 10^{-4}$ \cr
minimum surface brightness in fit ${S_{min}}^a$ &
\hfil $3 \times 10^{-4}$ \cr
% radius of broad beam temperature region  & \hfil $16^\prime \se 1.04 \mpc$
%%\cr
\noalign{\medskip}
\noalign{\hrule}\cr
\noalign{\medskip}}}\hfil}}

% \smallskip
\vskip -0.3 truecm
\hskip 3.0truecm $^a$ units : ROSAT $\cntflux$.

\bigskip
% \vfill\eject

\centerline{\bf Table 3}
\smallskip
\centerline { Mean Mach Numbers within $\rfiveh$ }
\medskip
\vbox{\hbox to \hsize{\hfil\vbox{\halign {#\hfil&&\quad\hfil#\hfil\cr
\noalign{\hrule}\cr
\noalign{\smallskip}
\noalign{\hrule}\cr
\noalign{\medskip}
Sample & \hfil $\langle v_r \rangle/c_s$ & \hfil $\langle v_r^2
\rangle /c_s^2$ \cr
\noalign{\smallskip}
\noalign{\hrule}\cr
\noalign{\medskip}
  EJ & \hfill $0.001 \pm 0.016$  & \hfil  $0.041 \pm 0.010 $ \cr
  2F & \hfill $-0.022 \pm 0.022$   & \hfil $0.069 \pm 0.018 $ \cr
 NFW & \hfill $-0.080 \pm 0.119$  & \hfil $0.261 \pm  0.077 $ \cr
 Eds & \hfill $-0.008 \pm 0.033$  & \hfil $0.112 \pm 0.019 $ \cr
 Fl2 & \hfill $-0.012 \pm 0.004$  & \hfil $0.022 \pm 0.009$ \cr
 Op2 & \hfill $-0.005 \pm 0.014$  & \hfil $0.045 \pm 0.020$ \cr
\noalign{\smallskip}
\noalign{\hrule}\cr
\noalign{\medskip}}}\hfil}}

\vfill\eject

\centerline{\bf Table 4}
\smallskip
\centerline { $\beta$-Model Accuracy}
\medskip
\vbox{\hbox to \hsize{\hfil\vbox{\halign {#\hfil&&\quad\hfil#\hfil\cr
\noalign{\hrule}\cr
\noalign{\smallskip}
\noalign{\hrule}\cr
\noalign{\medskip}
% \noalign{\hskip 3.0truecm $\Omega \se 1$~~~~~~$\Omega_0 \se 0.2$}\cr
\hfil & ~~~~ $\Omega \se 1$ & & ~~~~ $\Omega_0 \se 0.2$ & \cr
$\delta_c^{est}$ & ~~ ${\bar {\cal X}}^a$ \hfil &
${\sigma_{\cal X}}$ \hfil & ~~ ${\bar {\cal X}}^a$ &
${\sigma_{\cal X}}$ \hfil \cr
\noalign{\smallskip}
\noalign{\hrule}\cr
\noalign{\medskip}
  100  & ~~ 1.46 \hfil & 0.53 \hfil & ~~ 2.08 \hfil & 0.50 \hfil \cr
  250  & ~~ 1.15 \hfil & 0.36 \hfil & ~~ 1.60 \hfil & 0.33 \hfil \cr
  500  & ~~ 1.02 \hfil & 0.29 \hfil & ~~ 1.34 \hfil & 0.24 \hfil \cr
  1000  & ~~ 0.94 \hfil & 0.23 \hfil & ~~ 1.15 \hfil & 0.19 \hfil \cr
  2500  & ~~ 0.87 \hfil & 0.16 \hfil & ~~ 1.00 \hfil & 0.14 \hfil \cr
\noalign{\smallskip}
\noalign{\hrule}\cr
\noalign{\medskip}}}\hfil}}

\smallskip
{\obeylines \parskip=0pt
\hskip 3.5truecm $^a$ ${\cal X} \se M^{est}/M^{true}$.
}

\bigskip
\bigskip

\centerline{\bf Table 5}
\smallskip
\centerline { Scaling Law Accuracy}
\medskip
\vbox{\hbox to \hsize{\hfil\vbox{\halign {#\hfil&&\quad\hfil#\hfil\cr
\noalign{\hrule}\cr
\noalign{\smallskip}
\noalign{\hrule}\cr
\noalign{\medskip}
% \noalign{\hskip 3.0truecm $\Omega \se 1$~~~~~~$\Omega_0 \se 0.2$}\cr
\hfil & ~~~~ $\Omega \se 1$ & & ~~~~ $\Omega_0 \se 0.2$ & \cr
$\delta_c$ & ~~ ${r_{10}(\delta_c)}^a$ \hfil & ${\sigma_{\cal X}}$ \hfil &
{}~ ${r_{10}(\delta_c)}^a$ \hfil & ${\sigma_{\cal X}}$ \hfil \cr
\noalign{\smallskip}
\noalign{\hrule}\cr
\noalign{\medskip}
  100  & ~~ 4.89 \hfil & 0.20 \hfil & ~~ 4.78 \hfil & 0.20 \hfil \cr
  250  & ~~ 3.37 \hfil & 0.18 \hfil & ~~ 3.31 \hfil & 0.16 \hfil \cr
  500  & ~~ 2.48 \hfil & 0.15 \hfil & ~~ 2.48 \hfil & 0.14 \hfil \cr
  1000  & ~~ 1.79 \hfil & 0.11 \hfil & ~~ 1.87 \hfil & 0.12 \hfil \cr
  2500  & ~~ 1.11 \hfil & 0.08 \hfil & ~~ 1.25 \hfil & 0.10 \hfil \cr
\noalign{\smallskip}
\noalign{\hrule}\cr
\noalign{\medskip}}}\hfil}}

\smallskip
{\obeylines \parskip=0pt
\hskip 3.5truecm $^a$ Defined by equation (8), units are Mpc ($h \se 0.5$).
}

\vfill\eject

% \input refs.tex

% \magnification=1200
% \parindent=0.0truept
% \parskip=6.0truept
% \baselineskip=24.0 truept		% (APJ)
% \baselineskip=16.0 truept
% \def \bref{\par \noindent \hangindent=1.5 truecm \hangafter=1}  %
							        %References
% \def \etal      {{\it et al.\ }}

\centerline {\bf References}
\bigskip

Bartelmann, M., 1995, \aa, 299, 11.

Bartelmann, M., Steinmetz, M. \& Weiss, A. 1995, \aa, 297, 1.

Bartlett, J.G., Blanchard, A., Silk, J. 1995, Science,
267, 980.

Bertschinger, E. 1985, \apjs 58, 39.

Bertschinger, E. 1987, \apj, 323, L103.

Bird, C.M. 1995, \apj, 445, L81.

Briel, U.G. \& Henry, J.P. 1994, \nature, 372, 439.

Briel, U.G., Henry, J.P., \& B\"ohringer, H. 1992, \aa, 259, L31.

\bref
Briel, U.G., Henry, J.P., Schwarz, R.A.,  B\"ohringer, H. \& Ebeling,
H., Edge, A.C., Hartner, G.D., Schindler, S., Tr\"umper, J. \& Voges,
W. 1991, \aa, 246, L10.

Bonnet, H., Mellier, Y. \& Fort, B. 1994, \apj, 427, L83.

Buote, D.A. \& Tsai, J. C. 1995, \apj, in press.

Carr, B. 1993, Nature, 363, 16.

Cavaliere, A. \& Fusco--Femiano, R. 1976, \aa, 49, 137.

Cen, R. \& Ostriker, J.P. 1992, \apj, 399, L113.

\bref
Copi, C.J., Schramm, D.N. \& Turner, M.S. 1995, Science, 267, 192.

Crone, M.M., Evrard, A.E. \& Richstone, D.O. 1994, \apj, 434, 564.

Crone, M.M., Evrard, A.E. \& Richstone, D.O. 1995, \apj, submitted.

David, L.P., Jones, C. \& Forman, W. 1995, \apj, 445, 578.

Dressler, A. \& Shectman, S.A. 1988, \aj, 95, 985.

\bref
Durret, F., Gerbal, D., Lachi\`eze-Ray, G., Lima-Neto, G. \& Sadat,
R. 1994, \aa, 287, 733.

Evrard, A.E. 1990, \apj, 363, 349.

Evrard, A.E., Mohr, J.J., Fabricant, D.G. \& Geller, M.J. 1993, \apj,
419, L9.

Fahlman, G., Kaiser, N., Squires, G. \& Woods, D. 1994, \apj, 437, 56.

Freedman, W. \etal 1994, \nature, 371, 757.

Gunn, J.E., \& Gott, J.R. 1972, \apj, 209, 1

\bref
Hata, N., Scherrer, R.J., Steigman, G., Thomas, D. \& Walker,
T.P. 1994, OSU-TA-26/94.

Hoffman, Y. 1988, \apj, 328, 489.

Hughes, J. P. 1989, \apj, 337, 21.

Ikebe, Y. \etal 1994, in {\it Clusters of Galaxies },
eds. F. Durret, A. Mazure \& J. Tranh Thanh Van,
(Gif--sur--Yvette: Editions Frontieres), p. 167.

Krauss, L. \& Kernan, P.J. 1994, \apj, 432, L79.

Loeb, A. \& Mao, S. 1994, \apj, 435, L109.

Miralde-Escud\'e, J. \& Babul, A. 1995, \apj, 449, 18.

Miyaji, T. \etal 1993, \apj, 419, 66.

Metzler, C.A. \& Evrard, A.E. 1994, \apj, 437, 564.

Metzler, C.A. \& Evrard, A.E. 1995, in preparation.

\bref
Mohr, J.J., Evrard, A.E., Fabricant, D.G., \& Geller, M.J. 1995, \apj,
447, 8.

Mould, J. \etal 1995, \apj, 449, 413.

Mushotzky, R. 1994,  in {\it Clusters of Galaxies },
eds. F. Durret, A. Mazure \& J. Tranh Thanh Van,
(Gif--sur--Yvette: Editions Frontieres), p. 167.

Navarro, J.F., Frenk, C.S. \& White, S.D.M. 1995a, \mnras, 275, 720.

Navarro, J.F., Frenk, C.S. \& White, S.D.M. 1995b, \apj, submitted.

Navarro, J.F. \& White, S.D.M., (1993) \mnras, 265, 271.

Navarro, J.F. \& White, S.D.M., (1994) \mnras, 267, 401.

Neumann, D.M. \& B\"ohringer, H. 1995, \aa, in press.

Pearce, F., Thomas, P.A. \& Couchman, H.M.P. (1994) \mnras, 268, 953.

Pildis, R.A., Bregman, J.N. \& Evrard, A.E. 1995, \apj, 443, 514.

\bref
Ponman, T.J., Allan, D.J., Jones, L.R., Merrifield, M., McHardy,
I.M., Lehto, H.J., \& Luppino, G.A. 1994, Nature, 369, 462.

\bref
Rottgering, H., Snellen, I., Miley, G., DeJong, J.P., Hanisch, R.J. \&
Perley, R. 1994, \apj, 436, 654.

Sasselov, D. \& Goldwirth, D. 1995, \apj, 444, 5.

Schindler, S. 1995, \aa, in press.

Schindler, S. \& B\"ohringer, H. 1993, \aa, 269, 83.

Schindler, S. \& M\"uller 1993, \aa, 272, 137.

\bref
Smail, I., Ellis, R.S., Fitchett, M.J. \& Edge, A.C. 1995, \mnras,
273, 277.

\bref
Squires, G., Kaiser, N., Babul, A., Fahlman, G., Woods, D., Neummann,
D. \& B\"ohringer, H. 1995, preprint (astro-pg/9507008).

\bref
Steigman, G. 1995, in {\it The Light Element Abundances}, ed. P. Crane
(Berlin: Springer), in press.

Tsai, J.C., Katz, N. \& Bertschinger, E. 1994, \apj, 423, 553.

Tyson, J.A., Valdes, F. \& Wenk, R.A. 1990, \apj, 349, L1.

\bref
Walker, T.P., Steigman, G., Schramm, D. N., Olive, K. A., \& Kang, H.
1991, \apj, 376, 51.

\bref
Watt, M. P., Ponman, T. J., Bertram, D., Eyles, C. J., Skinner, G. K.,
\& Willmore, A.P. 1992, \mnras, 258, 738.

White, D.A. \& Fabian, A.C. 1995, \mnras, 273, 72.

\bref
White, S. D. M., Navarro, J. F., Evrard, A. E., \& Frenk, C. S. 1993,
\nature, 366, 429.

\vfill\eject
\noindent{\bf Figure Captions}
\medskip
\parindent=0pt

{\bf Figure 1.} Synthetic ROSAT \xray surface brightness (left
column), emission weighted temperature (center column) and emission
weighted, projected velocity field (right column) for three orthogonal
projections of a single cluster from the EdS sample.  The cluster is
at an assumed redshift $z \se 0.04$ and bars in the figure show a
$10^\prime$ angular scale.  In the grey scale contours, the dark (or
light) bands are logarithmically spaced by factors of $10^{0.6}$ in
surface brightness, $10^{0.2}$ in temperature, with the third from
minimum dark band representing $10^{-3} \cntflux$ and $10^7 \, {\rm
K}$, respectively.  The velocity vectors are spaced every $2^\prime$,
and scaled such that $1^\prime \se 1000 \kms$.

{\bf Figure 2.}  Projected azimuthally averaged \xray surface brightness
(top) and emission weighted temperature (bottom) for the cluster shown
in Figure~1.  The top to bottom lines correspond to the top to bottom
projections, with data for the middle and bottom projections displaced
by factors of 10 (2) and 100 (4) for the surface brightness
(temperature) with respect to the top projection.
% Error bars derived
% from Poisson statistics assuming an exposure time of 7200s with the
% ROSAT PSPC are shown for the surface brightness.

{\bf Figure 3.} Radial Mach numbers $\langle v_r \rangle/c_s$ (left)
and $\langle v_r^2 \rangle/c_s^2$ (right) derived from the gas sound
speed $c_s^2(r) \se 5 kT(r)/3\mu m_p$ and the first and second moments
of the radial velocity field for all the runs in the ensemble.  All
quantities are local values measured in radial shells.  The
overdensity $\delta_c$ is used as a radial coordinate; note the inverted
axis.  The dashed line shows $\delta_c \se 500$, our conservative
estimate for the boundary of the hydrostatic region.

{\bf Figure 4.} Scaling between cluster size, as measured by
$\rfiveh$, and emission weighted temperature for all the models.
Symbol types correspond to different models, as shown in the legend.
The data are well fit by equation (5).  Each model
appears three times, from three orthogonal projections.

{\bf Figure 5.} Three dimensional temperature profiles for all the
clusters in the ensemble.  The temperature in radial bins is expressed
in terms of the average, emission weighted temperature $T_{\rm X}$ and
radius is normalized to the cluster size $\rfiveh$.

{\bf Figure 6.} Accuracy of the $\beta$-model mass estimates
as a function of the
estimated density contrast $\delta_c^{est}$, equation (6), for the
ensemble.  Each model appears three times, from orthogonal
projections.

{\bf Figure 7.} Histograms of the estimated-to-true mass ratios
derived from the $\beta$-model evaluated at
three different estimated density contrasts.  Dashed
vertical lines show an error of $\pm 40\%$.

{\bf Figure 8.} \xray surface brightness maps of the six worst
underestimates from the $\Omega \se 1$ ensemble.  Values of the
estimated-to-true mass ratio are shown above each panel.  Within each
panel, the light and bold circles represent the true and estimated
values of $\rfiveh$, respectively.  Strongly bimodal or complex images
usually result in poor $\beta$-model mass estimates.

{\bf Figure 9.} Histogram of the estimated-to-true mass ratios from
the $\beta$-model for the $\Omega \se 1$ combined sample at
$\delta_c^{est} \se 500$.

{\bf Figure 10.} Histogram of the estimated-to-true mass ratios from
the scaling law method for the $\Omega \se 1$ combined sample at
$\delta_c \se 500$.  The distribution is unbiased by construction.

{\bf Figure 11.} Scatter plot of the deviations in $\beta$ and $T_{\rm
X}$ (defined in the text) for all the models in the ensemble.  Point
styles are the same as those used in Figure~4.

\vfill\eject
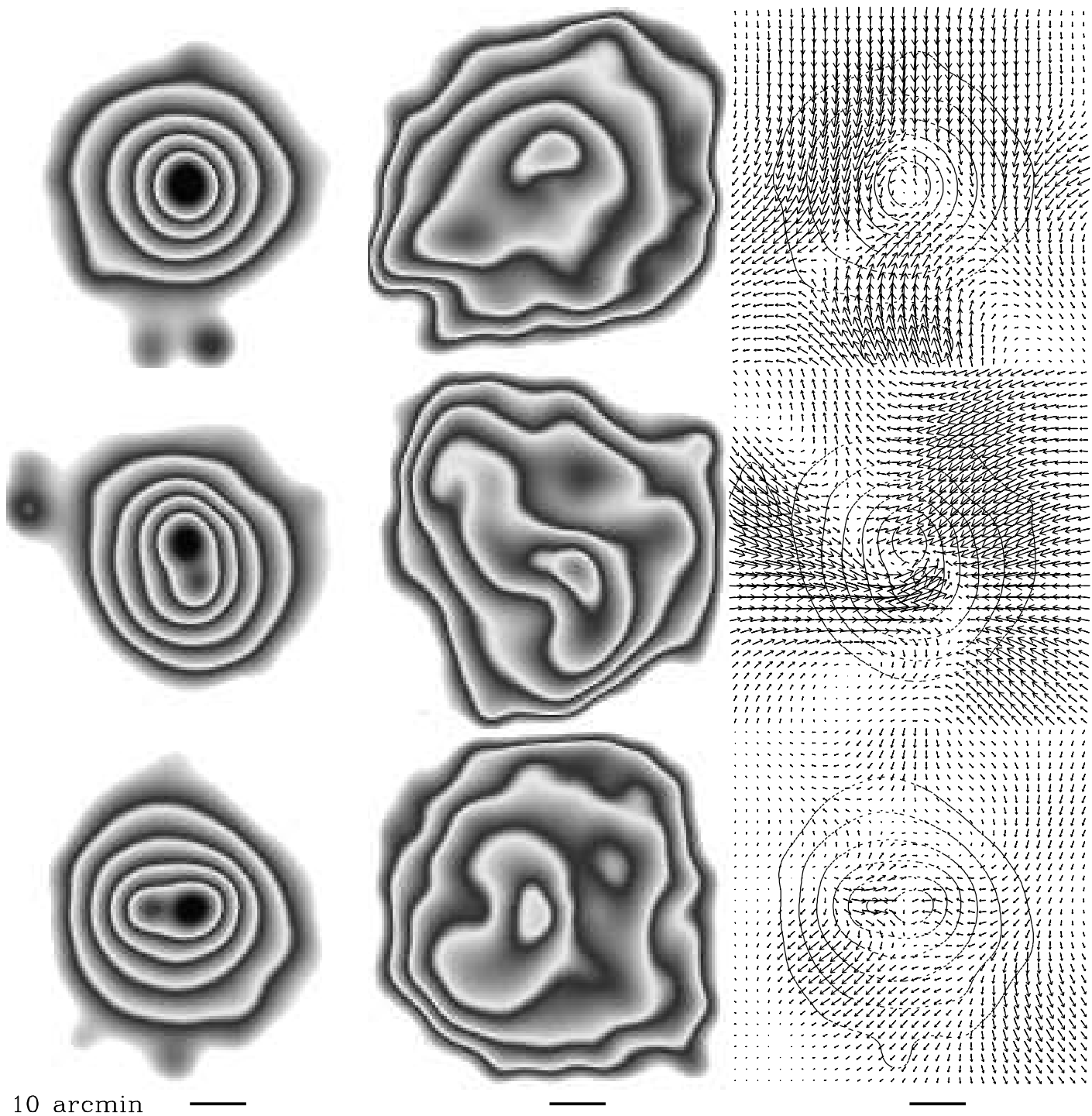
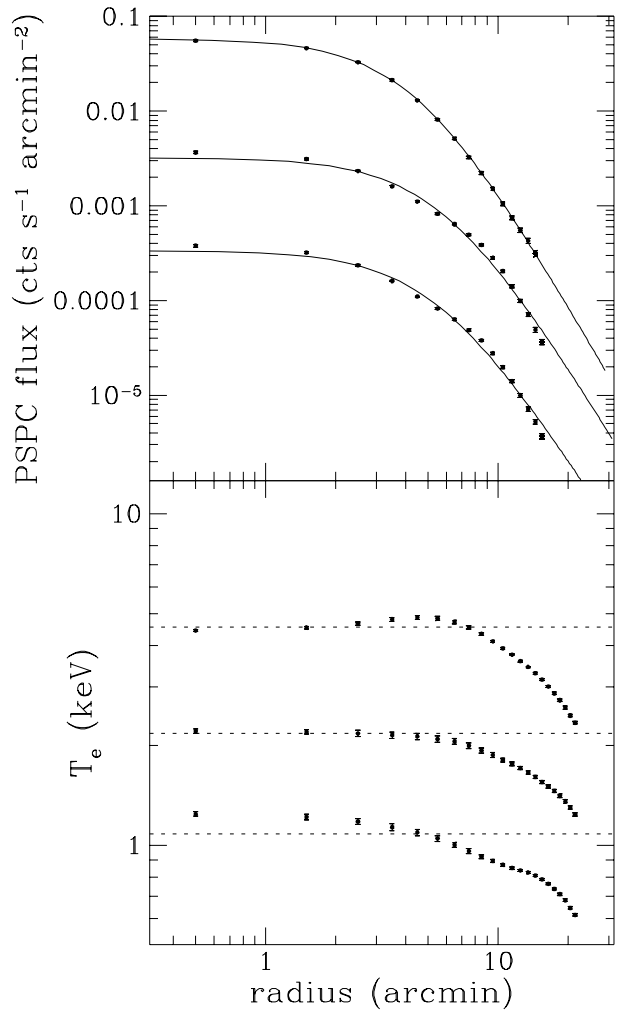
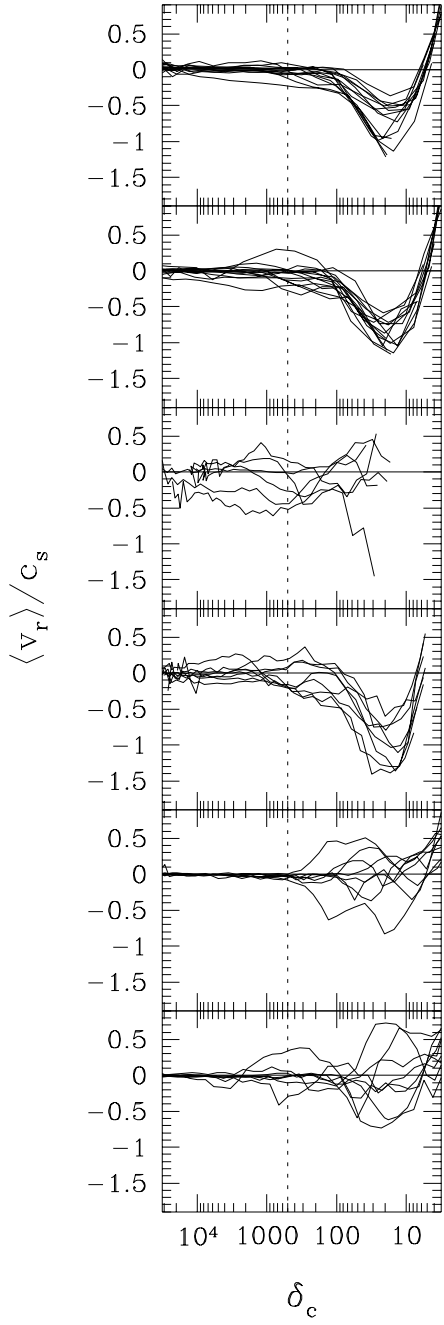
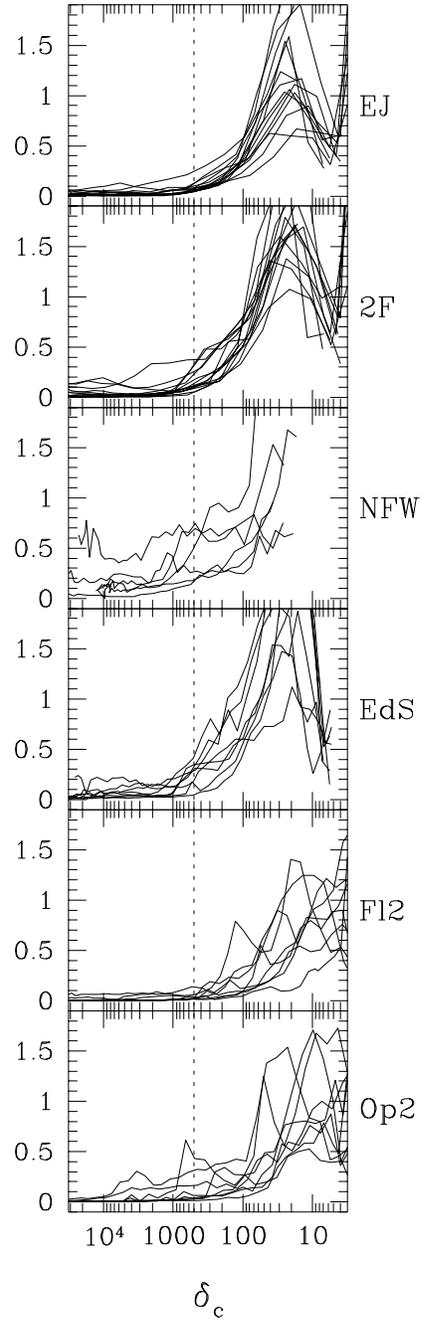
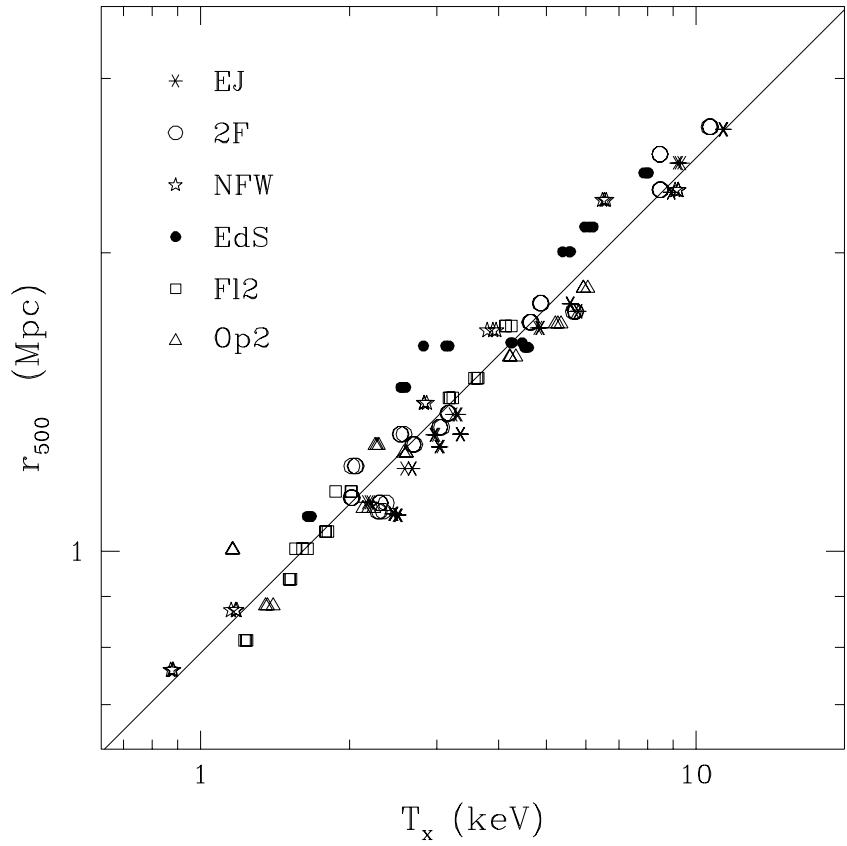
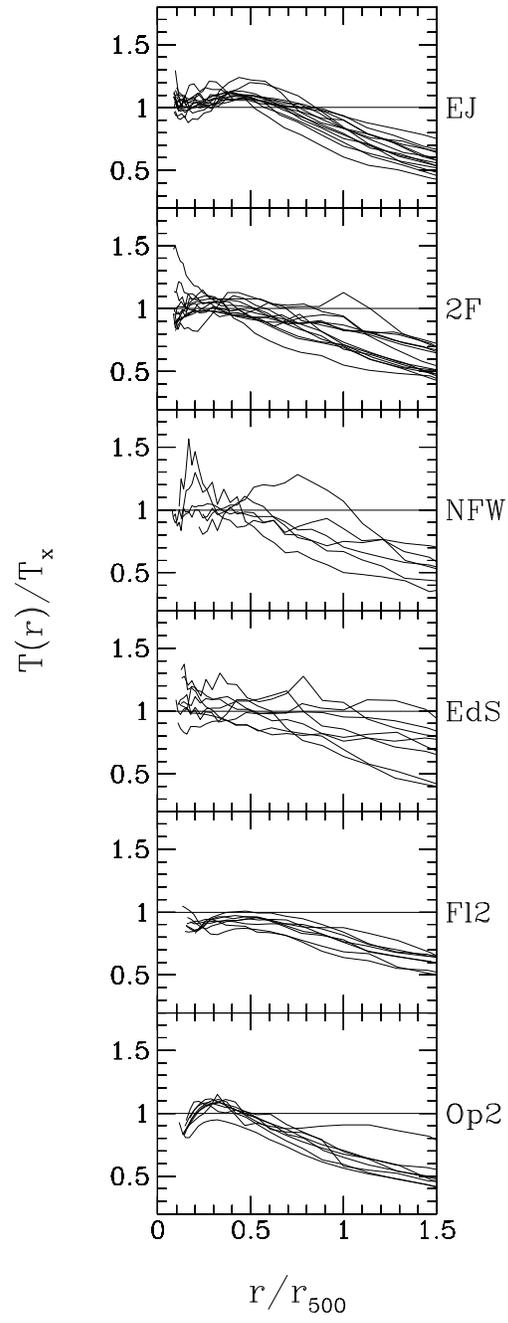
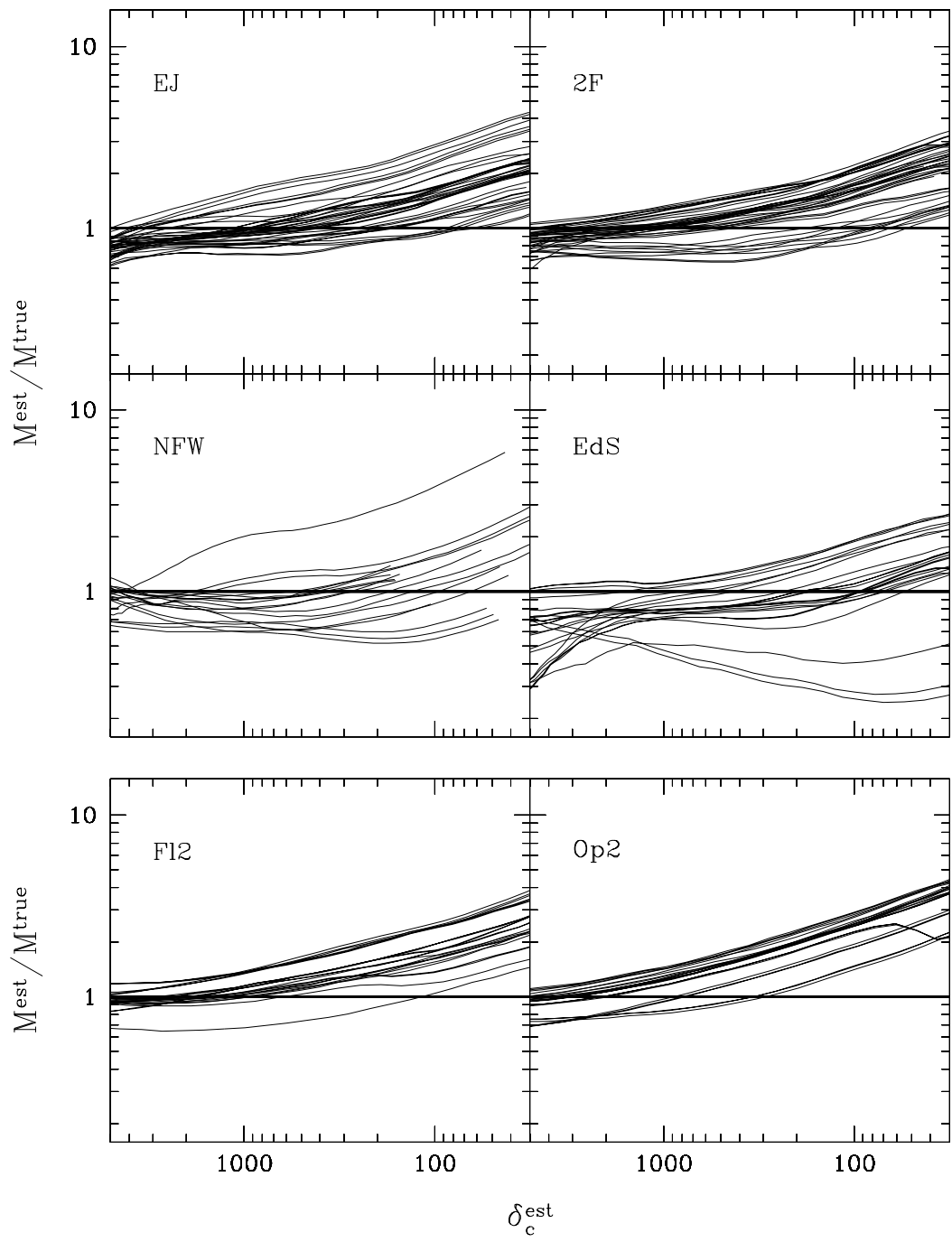
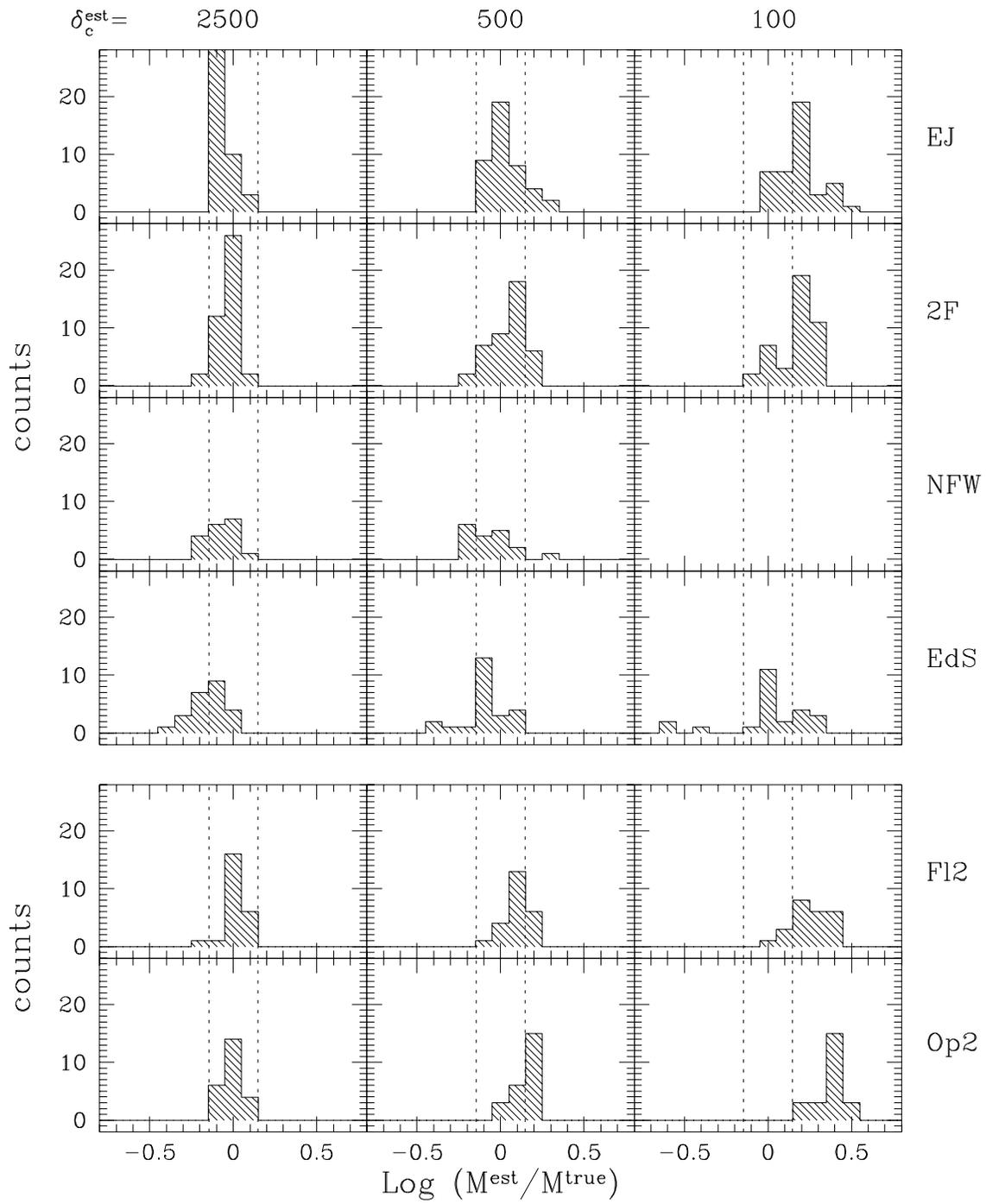
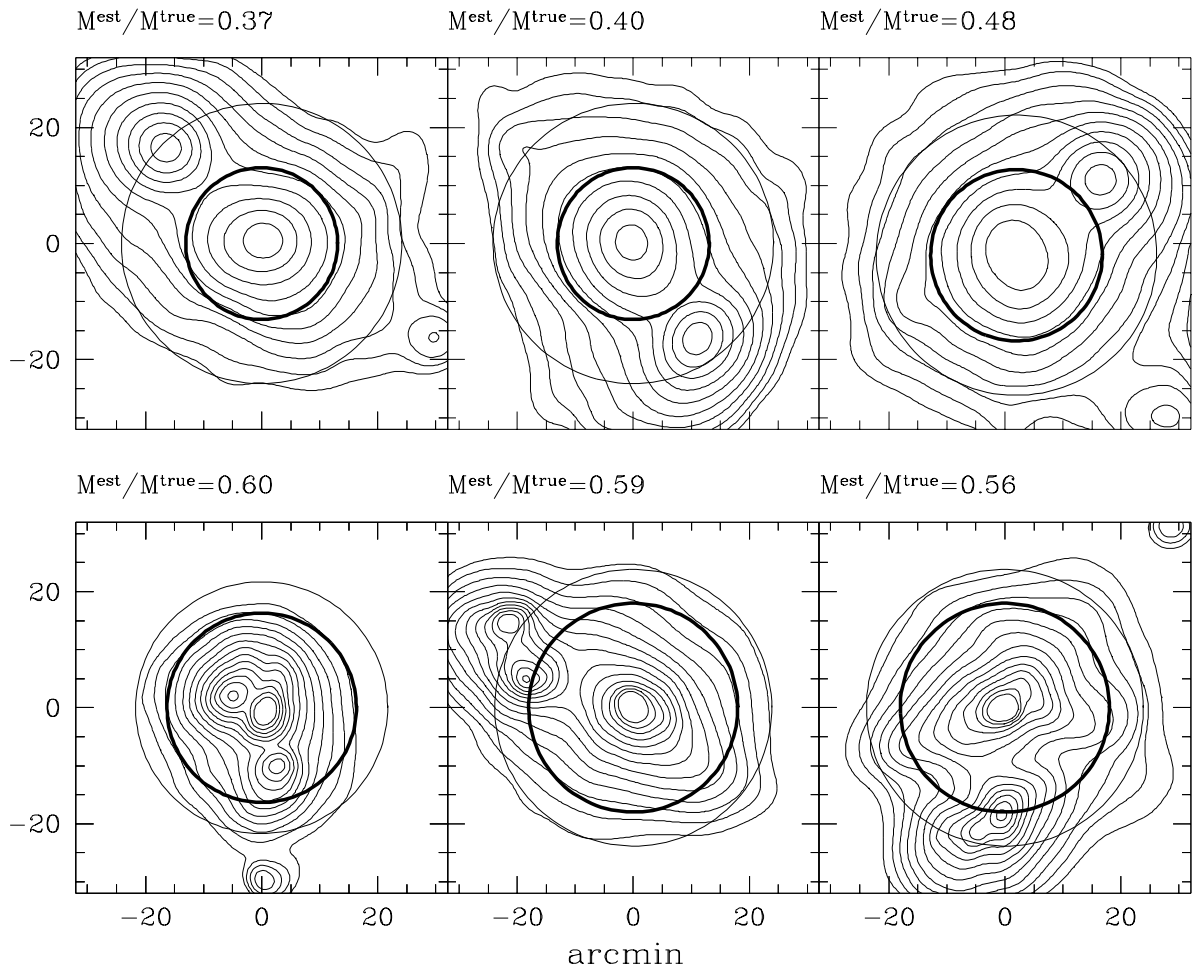
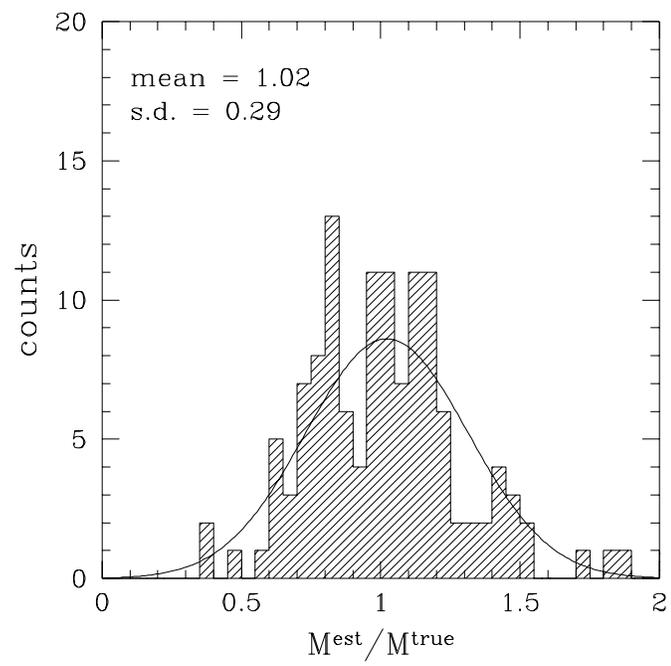
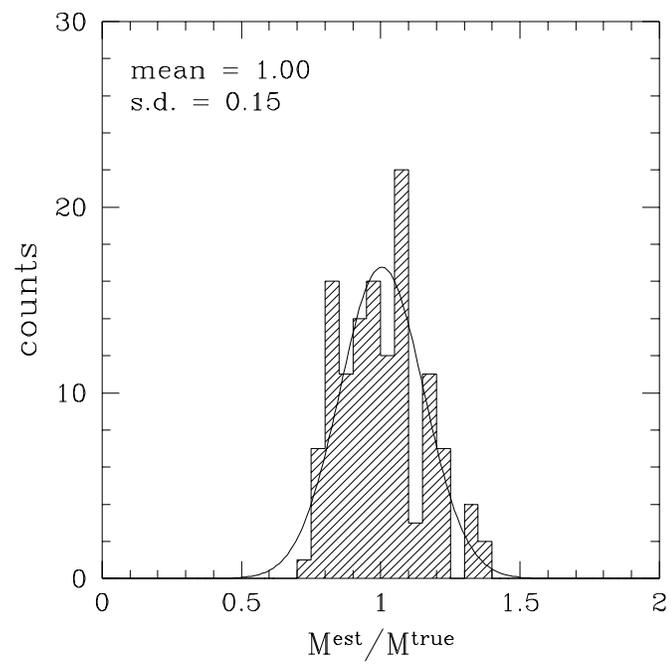
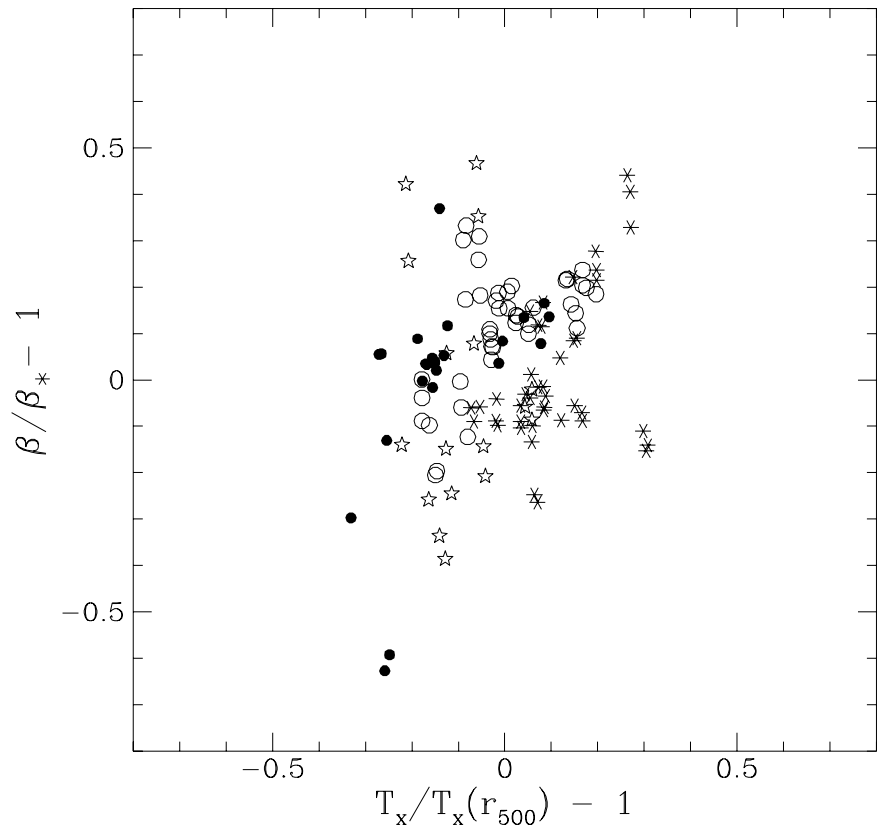
\end